\documentclass[prb,twocolumn,amsmath,amssymb]{revtex4-1}
\usepackage{graphicx}
\usepackage{bm}
\usepackage{dcolumn}
\usepackage{subfigure}
\usepackage[usenames]{color}

\begin{document}
\title{ Connectivity of the Icosahedral Network and a Dramatically Growing Static Length Scale in Cu-Zr Binary Metallic Glasses}
\author{Ryan Soklaski, Zohar Nussinov, Zachary Markow, K.F. Kelton, and Li Yang}

\affiliation{Department of Physics, Washington University in St.
Louis, St. Louis, MO 63130, USA}

\begin{abstract}
We report on and characterize, via molecular dynamics (MD)
studies, the evolution of the structure of Cu$_{50}$Zr$_{50}$ and
Cu$_{64}$Zr$_{36}$ metallic glasses (MGs) as temperature is
varied. Interestingly, a \textit{percolating icosahedral network}
appears in the Cu$_{64}$Zr$_{36}$ system as it is supercooled.
This leads us to introduce a static length scale, which grows
dramatically as this three dimensional system approaches the glass
transition. Amidst interpenetrating connections,
non-interpenetrating connections between icosahedra are shown to
become prevalent upon supercooling and to greatly enhance the
connectivity of the MG's icosahedral network. Additionally, we
characterize the chemical compositions of the icosahedral networks
and their components. These findings demonstrate the importance of
non-interpenetrating connections for facilitating extensive
structural networks in Cu-Zr MGs, which in turn drive dynamical
slowing in these materials.
\end{abstract}

\maketitle

\section{Introduction}

Understanding the structure-property relationships of metallic
glasses (MG) is a pressing matter in the field of material
science\cite{Greer07,Kelton03,Sillescu99,Appig06,Berthier07}. The
culmination of past research reveals that the vast diversity of MG
species, and their properties, are rooted in an equally-broad set
of structural archetypes\cite{MaReview,Miracle2004,Sheng2006}.
Perhaps even more cumbersome is a decades old outstanding problem
in the study of glasses-namely explaining the extreme slowing of
dynamics in a liquid as it is supercooled towards a glass
transition \cite{Montanari06,Tanaka10,Mosayebi10,
Berthier05,Karmakar10,Charb12,Biroli08,Bouchaud04,Coslovich12,Smarajit09,
Kurchan09,Ronhovde12,Lad12,Lubchenko07,Kirkpatrick89,Tarzia07,
Gotze99,Mayer06,Garrahan02,Nussinov04,Tarjus05,Ritort03,
Aharonov07, Berthier11,Chandler10,Hocky12}. Many theories, such as
the theory of random first order transitions
\cite{Lubchenko07,Kirkpatrick89}, spin glass
approaches\cite{Tarzia07}, and
others\cite{Nussinov04,Tarjus05,Hocky12} predict the existence of
rapidly growing length scales (in particular, a static length
scale) to accompany the marked increase of the viscosity of
supercooled liquids. In fact, there are rigorous results
predicting the appearance of such lengths \cite{Montanari06}.
While much progress\cite{Tanaka10,Smarajit09,Mosayebi10,
Berthier05,Karmakar10,Hocky12,Biroli08,Bouchaud04,Kurchan09,Coslovich12,Ronhovde12}
has been made in studying the structures of various glass-forming
systems, to date, no notable increase in standard static length
scales has been observed in most studies.

Cu-Zr is a popular representative of Early Transition Metal-Late
Transition Metal MGs; this is, in part, because it has a high
glass forming ability (GFA) for a broad range of compositions
\cite{Tang2004,PYu2005,OJKwon2006,YLi2008,Bendert2012}. The binary
composition of Cu-Zr reduces the complexity of the possible local
atomic structures, making this system ideal for the study of the
evolution of the spatial structure of liquids as they are
supercooled to form a glass. Recent research efforts demonstrate a
structural hierarchy within Cu-Zr MGs that appears to be central
to its structure-property
relationships\cite{Lad12,Schenk02,Cheng2008,Cheng2009,Nelson1983,Wakeda2010,Lee11}.
Strong evidence has been found to suggest that the presence of
Cu-centered full icosahedra is uniquely responsible for dynamical
slowing during the formation of the Cu-Zr
MGs\cite{Cheng2008,Lee11,Lad12}. Aside from their full and
distorted icosahedral Voronoi signatures, the Voronoi polyhedron
landscapes of Cu-Zr in the liquid and glass phases resemble one
another considerably\cite{Li2009}. The icosahedral clusters often
interpenetrate one another, so that five atoms coincide on two
icosahedral-shells. These interpenetrating connections of
icosahedra (ICOI) exhibit strong spatial
correlations\cite{Li2009,Lad12}
 and produce stable stringlike networks of icosahedra
\cite{Wakeda2010,Li2009,Tomida95,Shimono07,Takeuchi12,Lad12}.
Extensive ICOI clusters are found to posses a high
average elastic rigidity \cite{Wakeda2010} and to enhance more general
mechanical properties of Cu-Zr\cite{Lee11} MG.

In light of the importance of icosahedral clusters and ICOI in
shaping Cu-Zr MG, it is pressing to consider the roles of
non-interpenetrating connections in order to obtain a complete
picture of icosahedral networking in this amorphous alloy. To
date, the prominence of these connections has only been
noted\cite{Lad12}. In this paper, MD simulations are used to study
non-interpenetrating connections, and to characterize their effect
on the connectivity and composition of icosahedral networks in
Cu-Zr MGs, using Cu$_{50}$Zr$_{50}$ and Cu$_{64}$Zr$_{36}$ as
models. Our analysis reveals that, near the glass transition, a
large number of non-interpenetrating connections develop amongst
ICOI-structures and significantly enhance the unification of
icosahedral networks in these MGs, thus impacting their mechanical
properties. Considering the full icosahedral network affords us
the rather unique opportunity to identify, in Cu$_{64}$Zr$_{36}$,
\textit{a percolating icosahedral network that threads the entire
glassy system, providing a large scale static structure which
grows rapidly as the system approaches the glass transition}. We
also suggest that the non-interpenetrating connections are
particularly influential in MGs like Cu$_{50}$Zr$_{50}$, which
exhibits relatively few ICOI and yet is a good glass former.
Moreover, with all of the connection types being included, we
reveal that the icosahedral network and the otherwise liquidlike matrix\cite{Li2009} have
significantly different chemical compositions. For example, vertex
and face-sharing connections preferentially incorporate Zr atoms,
increasing the density of the icosahedral network. For both
alloys, the icosahedral network is Zr-rich, relative to the
overall sytem composition, and is immersed in a Cu-rich liquidlike
matrix.

In section II, we will proceed to discuss the methods for our
models of Cu-Zr, MD simulations, and structural analysis. Section
III characterizes the basic icosahedral ordering in our systems as
they are cooled from the liquid phase to the glass phase. In
section IV we measure the population and distribution of
connections in the icosahedral network. Sections V, VI, and VII,
respectively, presents weighted connectivities, introduces a
static length scale, and details the chemical compositions of the
icosahedral networks for Cu$_{50}$Zr$_{50}$ and
Cu$_{64}$Zr$_{36}$.

\section{Molecular Dynamics Simulations and Voronoi Tessellation}

The MD simulations conducted in this study use a semi-empirical
potential developed by Mendelev et al, which describes many-body
interactions using the Finnis and Sinclair (FS) generalization of
the embedded atom method (EAM)\cite{Mendelev09}. It was created,
in part, by being fitted to x-ray diffraction data, liquid
density, enthalpy of mixing data, and first-principles results.
The potential was developed with an emphasis on describing both
liquid and amorphous Cu-Zr alloys, and was shown to provide a very
good description of the structures and some properties of the
liquid and glass phases of Cu$_{64.5}$Zr$_{35.5}$. An earlier
version of this potential\cite{Mendelev07} was also shown to
produce reliable atomic structures for Cu$_{64.5}$Zr$_{35.5}$ MG
when compared to atomic structures generated by the reverse Monte
Carlo (RMC) method, which was constrained by independent x-ray
diffraction data and ab initio MD simulation results\cite{Li2009}.
The semi-empirical potential used in this paper accurately
recreates the properties of pure Zirconium and is expected to
yield reasonable structures for Cu$_{50}$Zr$_{50}$ as well.
Accordingly, the structure factor data and Voronoi polyhedra
distribution obtained for this composition are in good agreement
with the published x-ray and Voronoi polyhedra data of M. Li
\textit{et al}.\cite{Li2009}(not shown).

Canonical ($NVT$) systems of Cu$_{50}$Zr$_{50}$ and
Cu$_{64}$Zr$_{36}$ were simulated using LAMMPS\cite{Plimpton95}.
Each simulation consists of 2000 atoms contained in a box with
periodic boundaries. The initial configuration of an alloy is
generated randomly, and the alloy is allowed to melt and evolve
for .25 ns  at 1600K, using 5 fs time steps. The alloy is
subsequently quenched to its target temperature at a rate of
$10^{12}$ K/s, with the volume of the simulation box set to yield
zero pressure for this temperature. Finally, the quenched alloy
evolves for 1 ns at the target temperature, during which 78
snapshots of the atomic configuration are recorded. The target
temperatures used range from 300 K to 1600 K in steps of 100 K.
The process is repeated so that, ultimately, three simulations
with independent initial configurations were carried out for each
of the fourteen target temperatures. To check for system size
effects, canonical systems with 10,000 atoms and a target
temperature of 300 K were also simulated for both alloy
compositions. The approximate values of the glass transition
temperatures ($T_{g}$) for Cu$_{50}$Zr$_{50}$ and Cu$_{64}$Zr$_{36}$
were found to be $700$ K and $800$ K, respectively.

An analogous $NPT$ simulation ($P = 0$) of Cu$_{50}$Zr$_{50}$ with
a target temperature of 400 K was conducted for the sake of
comparison. The resulting average partial pair correlation
functions (ppcs) were found to be statistically identical to those
from the corresponding $NVT$ simulations.

The atomic structures in the MD-generated alloys were
characterized using Voronoi polyhedron-analysis for each
atom\cite{MaReview,Finney70,VoronoiG}. Using a procedure similar
to that of Hsu and Rahman, cluster vertices that were separated by
distances less than seven orders of magnitude of the simulation
box edge length were consolidated into one vertex\cite{Hsu79}.

\begin{figure}[t]
\includegraphics[width=0.46\textwidth]{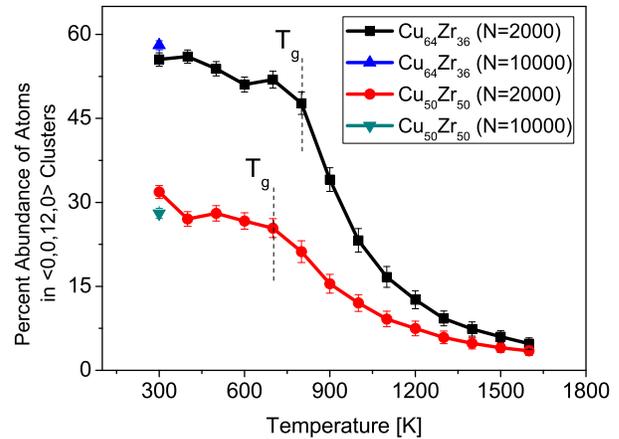}
\caption{(Color Online) Average percentage of atoms in a Cu-Zr
alloy that participate in a 13-atom icosahedral cluster (center
atom included). For $T =$ 300 K, data points are included for
a system size of $N = 10,000$ atoms in addition to $N = 2,000$
atoms. The error bars indicate the standard deviation of the
value. The glass transition temperatures
are marked for the respective compositions.} \label{fig:IcosOrder}
\end{figure}

\begin{figure*}[t]
\includegraphics*[scale=0.475]{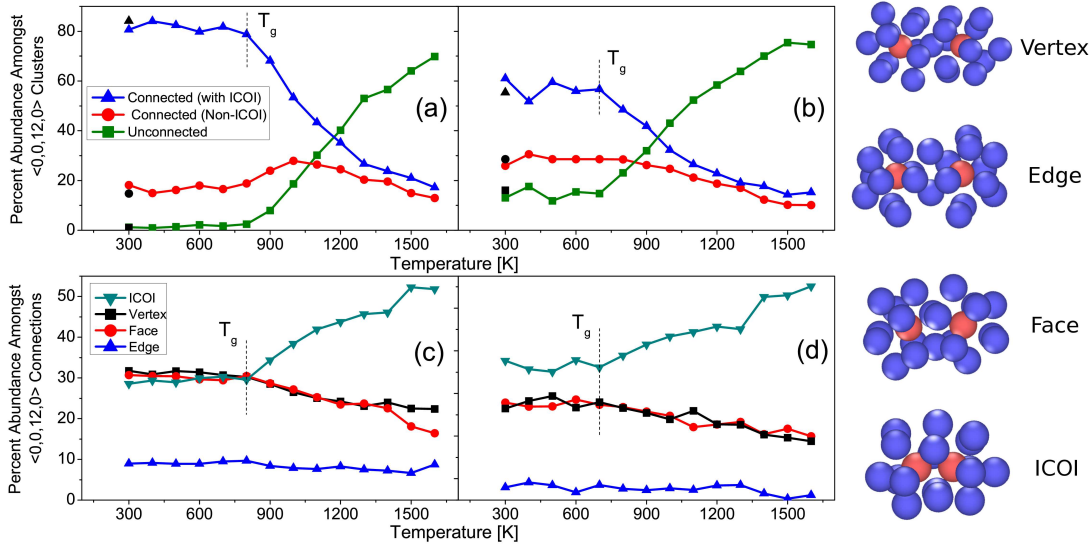}
\caption{(Color Online) [(a) and (b)] The average percentage of
$\langle$0,0,12,0$\rangle$ clusters involved in: at least one ICOI
connection, strictly non-ICOI connections, no connections for (a)
Cu$_{64}$Zr$_{36}$ and (b) Cu$_{50}$Zr$_{50}$. The black symbols
correspond to a system size of $N = 10,000$ atoms. [(c) and (d)]
The average percent distribution of connection types between
icosahedra in (c) Cu$_{64}$Zr$_{36}$ and (d) Cu$_{50}$Zr$_{50}$.
The cluster snapshots in the right column of the figure depict the
different types of connections between icosahedra. The red spheres
represent the center Cu-atoms of the two icosahedra involed.}
\label{fig:ConnAbun}
\end{figure*}

\section{Icosahedral Ordering in C\lowercase{u}-Z\lowercase{r} Alloys}

To study the development of icosahedral ordering in Cu-Zr, we
consider the results of our Voronoi tessellation analysis: a
Cu-atom with a Voronoi index $\langle$0,0,12,0$\rangle$
and its 12 Voronoi neighbors form a full icosahedral cluster.
The index $\langle$$n_{3},n_{4},...,n_{k},...$$\rangle$ specifies the number of
faces, $n_k$, with $k$ edges present on the Voronoi polyhedron. Zr-centered
full icosahedra are seldom observed amongst the Voronoi polyhedra
in these alloys, which is consistent with previous MD studies,
thus $\langle$0,0,12,0$\rangle$ clusters will henceforth refer
only to Cu-centered full icosahedra. Figure~\ref{fig:IcosOrder}
displays for a range of temperatures the percent-abundance of
atoms involved in $\langle$0,0,12,0$\rangle$ clusters. In their
liquid phases above 1300 K, fewer than 10\% of the atoms
participate in $\langle$0,0,12,0$\rangle$ clusters for both
compositions, but supercooling yields a great enhancement in the
degree of icosahedral ordering. Accordingly, roughly 28\% and 53\%
of the atoms of glassy Cu$_{50}$Zr$_{50}$ and Cu$_{64}$Zr$_{36}$,
respectively, participate in the structure of full icosahedra
(including their center Cu-atoms). This compositional trend of an
increase in icosahedral ordering, and subsequently an increase in
$T_{g}$, that accompanies an increase in the fractional
abundance of Cu-atoms has been demonstrated and studied
previously\cite{Cheng2008,Mattern2008}.

\begin{figure*}[t]
\includegraphics*[scale=0.44]{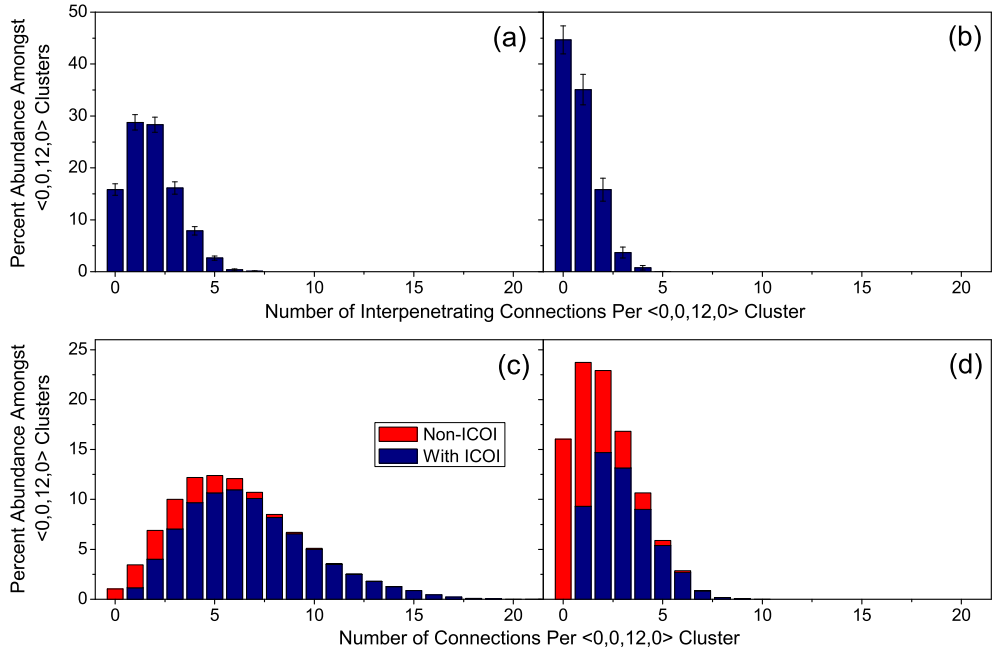}
\caption{(Color Online) [(a) and (b)] The average distribution of
the number of interpenetrating connections per icosahedron in (a)
Cu$_{64}$Zr$_{36}$ and (b) Cu$_{50}$Zr$_{50}$  MGs at 300 K ($N =
10,000$ atoms).  The error bars indicate the estimated error in
the mean value. [(c) and (d)] The average distribution of the
number of connections (vertex, edge, face-sharing, or
interpenetrating)  per icosahedron in (c) Cu$_{64}$Zr$_{36}$ and
(d) Cu$_{50}$Zr$_{50}$  MGs at 300 K ($N = 10,000$ atoms). These
are stacked-histogram plots in which the blue bars correspond to
icosahedra that participate in at least one interpenetrating
connection, and the red bars correspond to those that do not
participate in ICOI.} \label{fig:ConnDistr}
\end{figure*}

\section{Atomic Connections Between Icosahedra}

Below $T_{g}$, we observe that approximately 8.6\% of
Cu$_{64}$Zr$_{36}$'s Voronoi polyhedra in our simulation snapshots
are $\langle$0,0,12,0$\rangle$ clusters. If these clusters were
isolated from one another, it would require that 112\% of the
system's atoms participate in their structure. Considering that
only about 53\% of this alloy's atoms are actually involved in
these clusters, it is obvious that many clusters share atoms
amongst one another and are thus joined together. Therefore, two
$\langle$0,0,12,0$\rangle$ clusters are defined to be `connected'
if they share at least one atom. The inclusiveness of this
definition will be justified as we further study the details of
icosahedral structures. Icosahedra in our simulated Cu-Zr alloys
are observed sharing vertices, edges, and triangular faces in
addition to exhibiting the interpenetrating connections described
earlier. Examples of these connections are depicted in
Figure~\ref{fig:ConnAbun}. We will study the creation of these
connections between icosahedra during the process of quenching to
understand how they form the mechanically-stiff icosahedral
networks.

Figure~\ref{fig:ConnAbun} quantifies the development of
connections between icosahedra as the Cu-Zr alloys are quenched
from the liquid phase. Figures~\ref{fig:ConnAbun}a and
~\ref{fig:ConnAbun}b illustrate that as the number of icosahedra
in a Cu-Zr alloy increases rapidly during supercooling (refer to
Figure 1), the proportion of those participating in these stable
ICOI grow comparably. Above 1300K, less than a third of the
icosahedra are involved in ICOI for both alloys. The prevalence of
ICOI-based structures is heightened dramatically during
supercooling, and once Cu$_{64}$Zr$_{36}$ and Cu$_{50}$Zr$_{50}$
are arrested in their glass phases, approximately 80\% and 60\%,
respectively, of icosahedra have at least one interpenetrating
connection. Very few icosahedra are isolated in glassy Cu-Zr MGs,
e.g, about 1.7\% and 14\%, respectively, in Cu$_{64}$Zr$_{36}$
(Fig. 2a) and Cu$_{50}$Zr$_{50}$ (Fig.~\ref{fig:ConnAbun}b).
Figures~\ref{fig:ConnAbun}c and ~\ref{fig:ConnAbun}d display the
relative abundance of each connection-type in the alloys. They
show that vertex and face-sharing connections become increasingly
prevalent as the alloys are quenched and enter their glass phases.
For example, below $T_{g}$ in Cu$_{64}$Zr$_{36}$, these
connections are as common are ICOI (Fig.~\ref{fig:ConnAbun}c).
This is in stark contrast to the aforementioned results of
Figures~\ref{fig:ConnAbun}a and ~\ref{fig:ConnAbun}b, indicating
that, as the density of icosahedral clusters and the number of
ICOI increase during supercooling, vertex and face-sharing become
prevalent amidst ICOI-based structures. We will further study how
these non-interpenetrating connections are incorporated into the
icosahedral structures of Cu-Zr MG alloys.

Above 1300 K, the connections between icosahedra are constantly
forming and breaking (as are the icosahedra themselves) so that no
single connection between two icosahedra is observed in
consecutive snapshots (separated by 13 ps) of a simulation.
Viewing snapshots of the liquid alloys and assessing the average
abundances of the different connection-types thus provides a
measure of the relative prominences and stabilities of their
resulting structures. Figures~\ref{fig:ConnAbun}c and
~\ref{fig:ConnAbun}d show, for temperatures above 1000 K, that
ICOI comprise more than 40\%, a significant majority, of the
connections present in both alloys. This provides strong evidence
that the ICOI is in fact more robust than the other types of
connections and leads us to consider its role in the liquid alloys
as they cool. It is incorrect, however, to think of the glassy
icosahedral network as simple chains of interpenetrating
icosahedra, as our considerations of Figures~\ref{fig:ConnAbun}c
and ~\ref{fig:ConnAbun}d below $T_{g}$ demonstrate that these ICOI
structures also incorporate a significant number of
non-interpenetrating connections, as we elaborate on next.

\section{Non-Interpenetrating Connections and Network Connectivity}

\begin{figure}[t]
\includegraphics[width=0.40\textwidth]{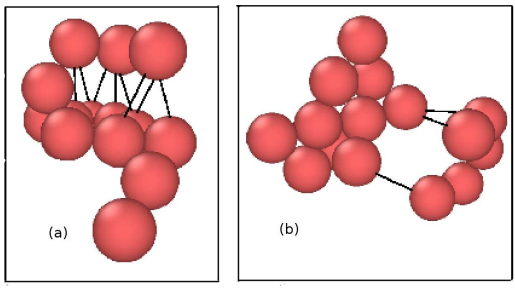}
\caption{(Color Online) Simulation snapshots\cite{OVITO} of
icosahedral structures containing a) pinning  and  b) bridging
connections. Only the central Cu atoms of the icosahedra are
displayed. Black lines are drawn between center Cu atoms to
indicate non-ICOI (vertex, edge, or face-sharing) connections
between their respective icosahedra.} \label{fig:PinBridge}
\end{figure}

To study the structures of the icosahedral networks found in Cu-Zr
MGs, we begin by characterizing the formative ICOI-structures.
Figures~\ref{fig:ConnDistr}a and~\ref{fig:ConnDistr}b show the
average distribution of the number interpenetrating connections
per icosahedron in the alloys at 300 K. As expected from our
considerations of Figure~\ref{fig:ConnAbun}, Cu$_{64}$Zr$_{36}$
contains more extensively connected ICOI-structures than does
Cu$_{50}$Zr$_{50}$ MG with an average of 2.2 interpenetrations per
connected-icosahedron versus 1.5, respectively. Both compositions
primarily exhibit chain-like and triangular arrangements of
interpenetrating icosahedra. These connectivity measurements are
corroborated and elaborated on by earlier
studies\cite{Lee11,Li2009}. Figures~\ref{fig:ConnDistr}c
and~\ref{fig:ConnDistr}d incorporate all connection types, and
provide a uniquely detailed characterization of the
icosahedral-structures found in these MGs. They display the
average distribution of the total number of connections per
icosahedron, while distinguishing icosahedra that participate in
ICOI from those that do not. From this perspective, the average
numbers of connections per interpenetrating icosahedron increase
to 6.8 and 3.0, respectively, for Cu$_{64}$Zr$_{36}$ and
Cu$_{50}$Zr$_{50}$. These distributions thus describe more
intricate, highly connected networks than do to those in
Figures~\ref{fig:ConnDistr}a and~\ref{fig:ConnDistr}b.
Furthermore, it becomes clear that interpenetrating icosahedra
frequently participate in other connection types and that they
serve as the nodes with the highest connectivity in the
icosahedral networks of the MGs. For example, in
Cu$_{64}$Zr$_{36}$ a sizable fraction of ICOI-icosahedra can be
found participating in ten or more connections, while
non-ICOI-icosahedra are seldom found with more that six
connections. It follows that significant number of
non-interpenetrating connections must occur within and between
ICOI-structures. These highly connected nodes in the networks
likely correspond to icosahedra with very low atomic mobilities,
which are known to be responsible for dynamical slowing in these
MGs\cite{Cheng2008,Lee11,Lad12}. When inspecting such densely packed
icosahedra, simulation snapshots reveal two common structures,
which are illustrated in Figure~\ref{fig:PinBridge}. In `pinned'
structures (Figure~\ref{fig:PinBridge}a), a single extended
ICOI-structure is folded on itself so that non-adjacent
(non-interpenetrating) icosahedra can be found to share vertices,
edges, and faces with one another, “pinning” the fold, and
producing a densely packed, stable, extended structure. A
`bridged' structure (Figure~\ref{fig:PinBridge}b) consists of
proximate ICOI-structures that are joined by non-interpenetrating
connections. The roles of these bridging connections are of
considerable interest since they serve to unify icosahedral
networks and thus potentially affect the structure-property
relationships of the MG. To demonstrate the significance of these
bridging connections, we measured the weighted connectivities of
the icosahedral networks.

\begin{figure}[h]
\includegraphics[width=0.4\textwidth]{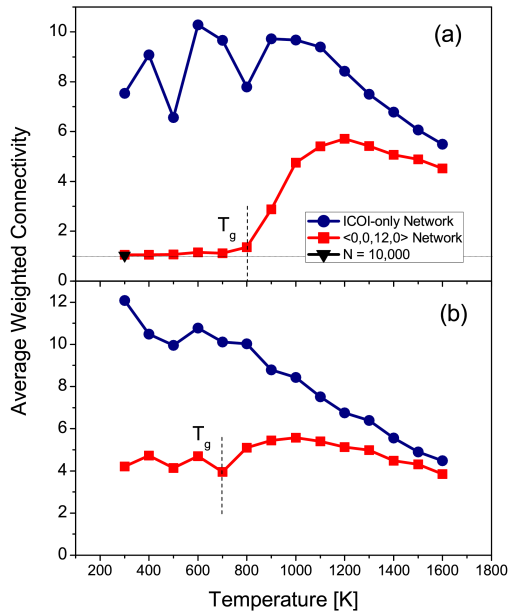}
\caption{(Color Online) The average weighted connectivity for the
ICOI-icosahedral network (blue circles) and the full icosahedral
network (red squares), which includes all connection types, for
(a) Cu$_{64}$Zr$_{36}$ and (b) Cu$_{50}$Zr$_{50}$}
\label{fig:Connectivity}
\end{figure}

\begin{figure}[b]
\includegraphics*[scale=0.3]{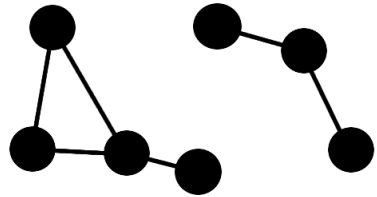}
\caption{An example graph, $G$, containing two connected subgraphs
of sizes 4 and 3, respectively. The weighted connectivity of $G$
is thus: $W_{G} = \frac{4}{4}+\frac{3}{4}$}
\label{fig:ExampleGraph}
\end{figure}

\begin{figure*}[t]
\includegraphics*[scale=0.38]{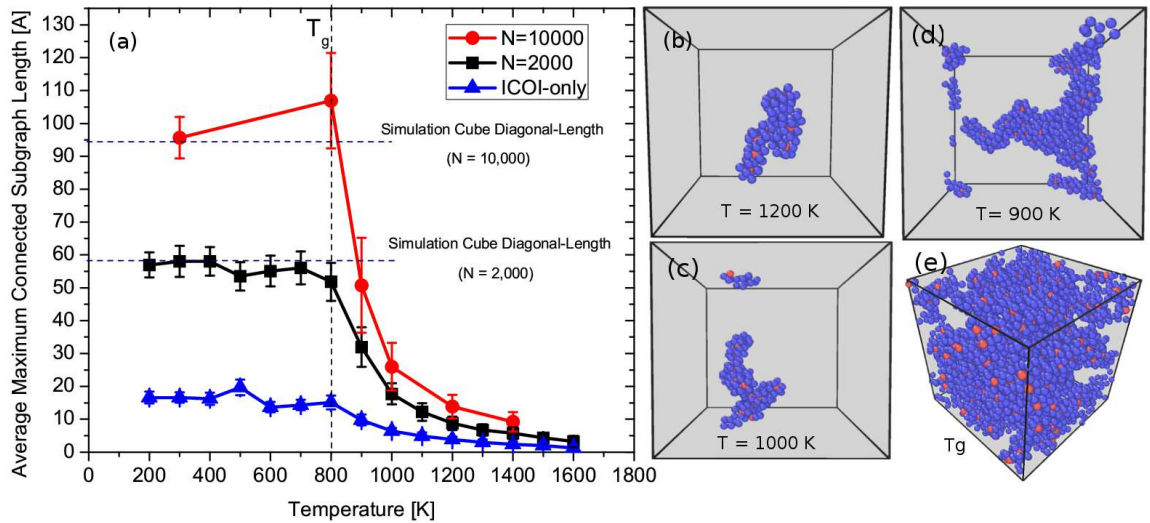}
\caption{(Color Online) (a) The average maximum
icosahedron-to-icosahedron distance within a connected subgraph of
the icosahedral network of Cu$_{64}$Zr$_{36}$. The simulation box
employs periodic boundaries, thus seemingly disjoint atoms are
attached to the connected subgraph across the box boundaries.
As indicated, the horizontal dashed lines denote $\sqrt{3}l$, where $l$ is the
respective simulation cube side length. The blue triangles correspond
to the longest connected subgraph lengths when only considering ICOI for
forming edges of the graph. The error bars indicate the standard
deviation in the maximum subgraph length. [(b) - (e)] Illustrative
simulation snapshots\cite{OVITO} depicting the longest connected
subgraph of the full icosahedral network at temperatures (b)
1200 K, (c) 1000 K, (d) 900 K, and (e) $T_{g}$ (800 K). The red
spheres depict center Cu atoms of $\langle$0,0,12,0$\rangle$
clusters. The blue spheres depict both Cu and Zr atoms that serve
 strictly as vertices of the $\langle$0,0,12,0$\rangle$ clusters.
The perspective in (e) was changed in order to show the extent of
the percolating subgraph.}
\label{fig:ClusterLength}
\end{figure*}

Consider the icosahedra of a Cu-Zr MG to be the nodes of a graph,
$G$, and their connections as the edges. That is, if two
icosahedra share a vertex, edge, or face, or if they
interpenetrate, then there is an edge between these two nodes. A
connected subgraph, $V$, of $G$ is defined to be a collection of
nodes and their edges such that any two nodes in $V$ are joined by
a path of edges and nodes, and that no edge connects to a node
outside of $V$. The size of $V$ is defined to be the number of
nodes in $V$. In this paper, an isolated node (one without edges)
will be considered a connected subgraph of size 1. Our graph, $G$,
is thus the union of all of its connected subgraphs. Let
\{$V_{i}$\}$_{i=1}^{n}$ be the set of $G$'s $n$ connected
subgraphs and \{$S_{i}$\}$_{i=1}^{n}$ be the corresponding set of
the sizes of the connected subgraphs. $S_{max}$ is the size of the
largest subgraph(s) ($S_{max} \equiv max(\{S_{i}$\}$_{i=1}^{n}$)).
The weighted connectivity of $G$ is then defined to be
\begin{equation}W_{G} \equiv \sum_{k=1}^n
\frac{S_{k}}{S_{max}}\end{equation} Thus a graph with $W_{G} = 1$
corresponds to a completely connected graph. A heuristic example
of a graph, its connected subgraphs, and its weighted connectivity
is provided in Figure~\ref{fig:ExampleGraph}.

\begin{figure*}[t]
\includegraphics*[scale=0.45]{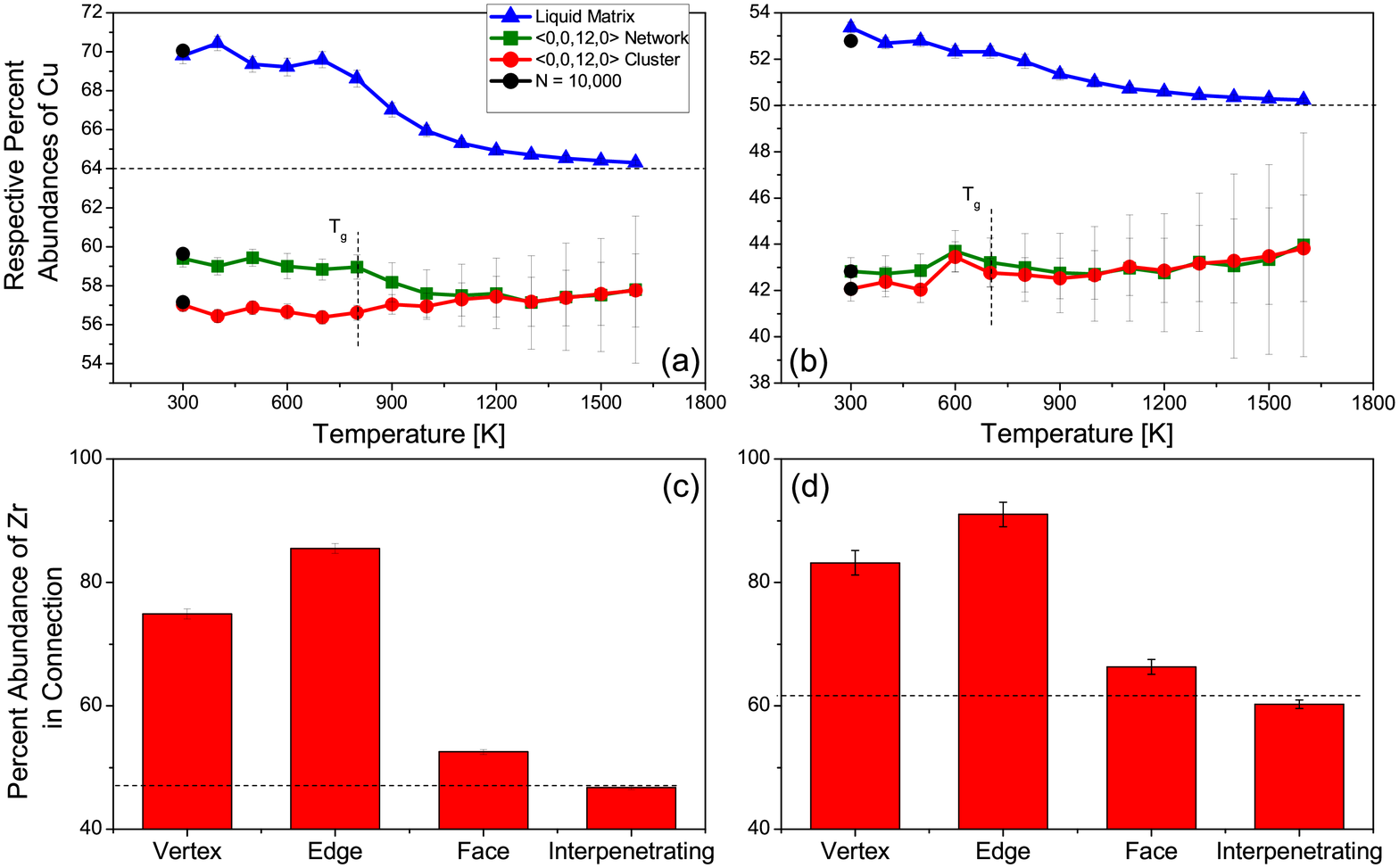}
\caption{(Color Online) [(a) and (b)] The average percent
composition of Cu present in: a single icosahedral cluster
(including the center atom), the full icosahedral network, the
liquid-like matrix (all atoms not involved in the icosahedral
network) for (a) Cu$_{64}$Zr$_{36}$ and (b) Cu$_{50}$Zr$_{50}$.
[(c) and (d)] Display the average percent-abundance of Zr found in
vertex, edge, face-sharing, and interpenetrating connections. The
dotted lines indicate the percent-composition of the shell of an
icosahedral cluster for (c) Cu$_{64}$Zr$_{36}$ and (d)
Cu$_{50}$Zr$_{50}$. The error bars indicate the error in the mean
values.} \label{fig:Composition}
\end{figure*}

Figures~\ref{fig:Connectivity}a and~\ref{fig:Connectivity}b show
the weighted connectivities of both the ICOI-networks and the
icosahedral networks that incorporate all connection types. It
should immediately be pointed out that the decrease in weighted
connectivity for high temperatures does not imply that the
networks are highly connected prior to supercooling. Rather, the
relatively low degree of icosahedral ordering at these
temperatures is responsible for the reduced measured
connectivities. In both compositions of Cu-Zr, the
non-interpenetrating connections that develop during supercooling
dramatically enhance the unification of the networks. The effect
of these connections is most obvious in Cu$_{64}$Zr$_{36}$ MG
(Fig.~\ref{fig:Connectivity}a); the weighted connectivity of the
icosahedral network decreases rapidly as the alloy is cooled from
1100 K down to 800 K ($T_{g}$). In contrast to this, the weighted
connectivity of the ICOI-network is not so significantly affected
by the supercooling process. This suggests that
non-interpenetrating connections are integral in linking the
disjoint pieces of the ICOI-network as the alloy approaches its
glass phase. In the case of Cu$_{64}$Zr$_{36}$, these bridging
connections lead to an almost completely unified icosahedral
network. The same network-unification was observed when the system
size was increased to $N = 10,000$ atoms. The weighted connectivity of
Cu$_{50}$Zr$_{50}$'s icosahedral network
(Fig.~\ref{fig:Connectivity}b) exhibits a similar enhancement in
comparison to its ICOI network, though network-unification is not
observed in this alloy.

These findings suggest that non-interpenetrating connections are
an important facet of icosahedral networking in Cu-Zr MGs and
perhaps in other MGs in which icosahedral ordering is prevalent.
The heavy presence of these connections amidst the most
highly-connected icosahedra suggests that they may assist in the
dynamical slowing of these clusters.  It is also reasonable to
expect that their roles in enhancing the density and structural
unification of the icosahedral network's MRO-structures contribute
significantly to the structural-mechanical properties that were
previously attributed only to
ICOI-structures\cite{Wakeda2010,Lee11}. Pins and bridges may be
particularly important for compositions of Cu-Zr MG for which the
development of ICOI is less extensive, and yet are good glass
formers. As indicated earlier, only about 60\% of the icosahedra
in Cu$_{50}$Zr$_{50}$ MG participate in interpenetrating
connections, but its icosahedral network is further developed via
other connection types (Fig.\ref{fig:ConnDistr}). We are also
interested in studying whether the icosahedral network's weighted
connectivity is simply related to the population density of
icosahedra, or if approaching a completely unified network reveals
more complex features.

\section{Static Length Scale}
Quenching a supercooled liquid into a glass produces major changes
in the system's dynamics, yet there is no obvious long range structural
change that accompanies this transition. Despite this, rigorous bounds have
been proven to exist between length and time scales in a glassy system
which mandate that a growing length scale must accompany the diverging relaxation
times of glass transitions\cite{Montanari06}. Recent research efforts
provide evidence for various growing correlation lengths
\cite{Lad12,Tanaka10,Mosayebi10,Berthier05,Karmakar10,Hocky12}. However, there are
findings that these scales (and in particular static length scales- i.e.,
those that can be ascertained from static snapshots of the system) do not
increase, when approaching the glass transition, rapidly enough, as
suggested by various theories , in order to account for the dramatic
increase in the relaxation times\cite{Charb12}. Static correlation lengths
have been studied via `point-to-set' correlations\cite{Biroli08,Bouchaud04,Hocky12},
three-point correlations \cite{Coslovich12}, pattern repetition
size\cite{Kurchan09}, and approaches employ graph theoretical tools
\cite{Ronhovde12} and shear rigidity penetration depths.
Currently, most of the intuitive length scales, especially static ones, do not
increase as drastically as does the relaxation time upon the glass transition.
As glasses are rigid, a length characterizing the scale on which the supercooled
liquid can support shear \cite{Ronhovde12} may be anticipated to naturally
increase as the system ``freezes'' into a rigid amorphous solid.

Motivated by these quests and, most notably, by the icosahedral
network-unification that we observe in Cu$_{64}$Zr$_{36}$ during its glass
transition, we introduce a static length scale, which interestingly exhibits
dramatic growth upon quenching. Our length scale is the longest
icosahedron-to-icosahedron distance within a connected subgraph of the full
icosahedral network. It may be conceivable that the percolating icosahedral
structure and the related increase in the interpenetrating icosahedral unit
structure that we find might be related to an increased rigidity of the system
as it rapidly cooled towards the low temperature glassy state. Connected
subgraphs that cross the periodic boundaries are appropriately `unfolded'
across the boundaries before the distance is measured, and the maximum length
is averaged across consecutive simulation snapshots.
Figure~\ref{fig:ClusterLength} depicts the rapid growth in the length scale as
Cu$_{64}$Zr$_{36}$ is cooled to $T_{g}$. From 1600 K down to $T_{g}$, the longest
ICOI-chain length grows by a factor of 13. For the full icosahedral networks
of the $N = 2,000$ atom and $N = 10,000$ atom systems, the longest connected subgraph lengths
increase by factors of 19 and 31, respectively. As expected from the percolating nature of
the network, and as indicated by the horizontal dashed lines in Fig.~\ref{fig:ClusterLength},
the extent by which the length scale grows is apparently limited only by the simulation cell size.

\section{Compositional Inhomogeneity}

Finally, we characterized the average chemical compositions of a
single icosahedral cluster (including its center atom), the
icosahedral network, and the different connection types. The
results are summarized in Figure~\ref{fig:Composition}. The
icosahedral networks of both Cu-Zr alloys were found to be Zr-rich
relative to their overall system compositions. This leaves the
remaining liquidlike matrix of the systems  to be Cu-rich, as
shown in Figures~\ref{fig:Composition}a
and~\ref{fig:Composition}b. Figure~\ref{fig:Composition}c displays
the average percent abundance of Zr in each type of connection.
Comparing these ratios to the average icosahedral-shell
composition (also displayed in Figure~\ref{fig:Composition}c)
reveals that non-interpenetrating connections do not appear to
occur randomly between icosahedra, rather, they preferentially
incorporate Zr-atoms. Because Zr has a larger atomic volume than
Cu, these connections increase the density of the icosahedral
network, while preserving the local composition of a single
icosahedral cluster.

\section{Conclusion}

In summary, we used a semi-empirical potential to conduct MD
simulations of Cu$_{50}$Zr$_{50}$ and Cu$_{64}$Zr$_{36}$ across
liquid and glass-phase temperatures. Our initial assessment of
icosahedral ordering and networking within these systems provides
additional support for the popular idea that interpenetrating
connections serve to create robust, formative icosahedral
structures. Non-interpenetrating connections are shown, however,
to become increasingly prevalent as the alloys approach their
glass phases. Viewing icosahedra as members of a network,
interpenetrating icosahedra frequently participate in
non-interpenetrating connections as well, and distinctly serve as
the nodes of the network with the most connections. In particular,
non-interpenetrating `pinning' and `bridging' connections are
prominent amongst ICOI-structures, and may be of particular
structural importance in Cu-Zr MGs in which ICOI are less
extensive. To further characterize the structure of the
icosahedral networks, we introduced the measure of weighted
connectivity and used it to show that the network of
Cu$_{64}$Zr$_{36}$ is almost completely unified, in contrast to
its ICOI-structures. Accordingly, the maximum connected-cluster
length lends itself as a static length scale, whose rapid growth
upon quenching is apparently limited only by the simulation cube
size. Lastly, the vertex, edge, and face-sharing connections are
shown to preferentially incorporate Zr atoms, amidst icosahedral
networks that are Zr-rich compared to the system compositions.

\section{Acknowledgments}

We thank Nick Mauro, and James Bendert for their useful
discussions. R.S. and L.Y. were partially supported by the
National Science Fundation (NSF) Grant No. DMR-1207141.
Z.N. was partially supported by NSF Grant No. DMR-1106293.
K.F.K was partially supported by NSF under grants DMR-08-56199
and DMR-12-06707, and NASA under grants NNX07AK27G and
NNX10AU19G.

\end{document}